%% file: main.tex
\documentclass[conference]{IEEEtran}
\IEEEoverridecommandlockouts
\usepackage{cite}
\usepackage{amsmath,amssymb,amsfonts}
\usepackage{algorithmic}
\usepackage[linesnumbered,ruled]{algorithm2e}

\SetCommentSty{commentfont}
\usepackage{graphicx, float}
\usepackage{xspace}
\usepackage{comment}
\usepackage{multirow}
\usepackage{enumitem}

\usepackage[most]{tcolorbox}

\usepackage{listings}
\usepackage{xcolor}
\usepackage{subfig}

\definecolor{codegreen}{rgb}{0,0.6,0}
\definecolor{codegray}{rgb}{0.5,0.5,0.5}
\definecolor{codepurple}{rgb}{0.58,0,0.82}
\definecolor{backcolour}{rgb}{0.95,0.95,0.92}

\lstdefinestyle{mystyle}{
    backgroundcolor=\color{backcolour},   
    language=Python,
    commentstyle=\color{codegreen},
    keywordstyle=\color{brown},
    numberstyle=\tiny\color{codegray},
    stringstyle=\color{codepurple},
    basicstyle=\ttfamily\footnotesize,
    breakatwhitespace=false,         
    breaklines=true,                 
    captionpos=b,                    
    keepspaces=true,                 
    numbers=left,                    
    numbersep=5pt,                  
    showspaces=false,                
    showstringspaces=false,
    showtabs=false,                  
    tabsize=2,
    moredelim=**[is][\color{red}]{@}{@}
}

\usepackage{makecell}
\def\BibTeX{{\rm B\kern-.05em{\sc i\kern-.025em b}\kern-.08em
		T\kern-.1667em\lower.7ex\hbox{E}\kern-.125emX}}
\begin{document}

\title{Empirical Study on Transformer-based Techniques for Software Engineering}

\author{\IEEEauthorblockN{Yan Xiao\IEEEauthorrefmark{1}, Xinyue Zuo\IEEEauthorrefmark{2}, Lei Xue\IEEEauthorrefmark{1}, Kailong Wang\IEEEauthorrefmark{3},  Jin Song Dong\IEEEauthorrefmark{2} and Ivan Beschastnikh\IEEEauthorrefmark{4}} \IEEEauthorblockA{\IEEEauthorrefmark{1}Shenzhen Campus of Sun Yat-sen University, China, 
		xiaoy367@mail.sysu.edu.cn, xuelei3@mail.sysu.edu.cn} 
	\IEEEauthorblockA{\IEEEauthorrefmark{2}National University of Singapore, Singapore, 
		e0376991@u.nus.edu, dcsdjs@nus.edu.sg}
	\IEEEauthorblockA{\IEEEauthorrefmark{3}Huazhong University of Science and Technology, China, Wangkl@hust.edu.cn}
	\IEEEauthorblockA{\IEEEauthorrefmark{4}Department of Computer Science, University of British Columbia, Vancouver, BC, Canada, 
		bestchai@cs.ubc.ca}}

\maketitle
\thispagestyle{plain}
\pagestyle{plain}


\begin{abstract}




Many Transformer-based pre-trained models for code have been developed
and applied to code-related tasks. In this paper, we review the
existing literature, examine the suitability of model architectures for different tasks, and look at the generalization ability of models on different datasets, and their resource consumption.

We examine three very representative pre-trained models for code:
CodeBERT, CodeGPT, and CodeT5, and conduct experiments on the top-4 most targeted software
engineering tasks that we found in our literature survey: Code Summarization, Bug Fixing, Bug Detection, and Code
Search. In our study, we showcase the capability of decoder-only models (CodeGPT) for specific generation tasks under state-of-the-art evaluation metrics and contest the common belief that the encoder-decoder architecture is optimal for general-purpose coding tasks. Additionally, we
found that the most frequently used models are not necessarily the
most suitable for certain applications and the developers' needs are not adequately addressed by current research. As well, we found that
the benchmark and frequent dataset for Bug Fixing and Code Summarization both fail to enable models to generalize onto other datasets for the same task (the frequent dataset refers to the dataset with the highest frequency used in literature other than the benchmark). We use statistical testing to support our conclusions from experiments. Finally, CodeBERT is highly efficient for understanding tasks, whereas CodeT5's efficiency for generation tasks is in doubt, as the highest resource consumption does not guarantee a consistent better performance on different metrics. We also discuss the
numerous practical issues in advancing future research on
transformer-based models for code-related tasks.


\end{abstract}

%



\input{intro}

\input{background}

\input{methodology}
\input{evaluation}

\input{related}
\input{conclusion}


\bibliographystyle{IEEEtran}
\bibliography{main}

%
%
%
%
%
%
%
%

\end{document}

%% file: intro.tex
\section{Introduction}
\label{sec:intro}

%
The availability of large natural language corpora and advances
in ML have led recent models to achieve extraordinary performance on
Natural Language Processing (NLP) tasks. Transformer-based
architectures~\cite{vaswani2017attention}, first introduced by Vaswani
et al. in 2017, are among the most successful model variants in this
field. Transformer-based models, like BERT (Bidirectional Encoder
Representations from Transformers) and GPT (Generative Pre-trained
Transformer), have revolutionized NLP tasks, including text
classification, sentiment analysis, and language generation.

Given the abundance of large software code corpora, Transformer-based models have also rapidly gained traction in software
engineering (SE)  research~\cite{le2022autopruner}, with hundreds of
transformer-related papers published in top-tier SE conferences and
journals in the past five years.
In many instances, these works have reported state-of-the-art performance
on a variety of SE tasks.
%
Some example applications of transformer-based techniques include
automated program
repair~\cite{fu2022vulrepair,zhang2022coditt5,xia2022less}, merge
conflict resolution~\cite{svyatkovskiy2022program,zhang2022using},
requirements
engineering~\cite{ezzini2022automated,anish2022automated,devine2022finding},
code and comment generation~\cite{chakraborty2022natgen,li2022auger},
code and machine
translation~\cite{chakraborty2022natgen,sun2022improving,he2022deepstl},
and more~\cite{le2022autopruner,huang2022prompt}. Model structures,
like encoder-only, decoder-only, and
encoder-decoder~\cite{chakraborty2022natgen}, together with the
different pre-training objectives, such as generative objectives and
denoising objectives, also add to the diversity of work in this space.


The excitement around these transformer-based models, however, must be
tempered with a careful assessment of their advantages and
pitfalls. This is the focus of this paper. 


%


In this paper, we take a step back and reflect on the copious amount
of work that has been published in this area thus far. We study 282
papers published at 27 top conferences and journals during 2017-2022.
We consider which models are being used in these papers, which SE
applications they target, benchmarks that they use, and other key
characteristics of this quickly growing body of work. We then closely
look at the performance of the top models from the literature on the
most popular applications and review the corresponding model
generalizability and computational complexity. Throughout, we frame our discussion in terms of
\emph{promises} and \emph{perils} to help position SE research that
relies on transformer-based models on a firmer footing.

The closest related empirical studies of this
rich research space have considered a fixed applications
set~\cite{lu2021codexglue,zeng2022extensive,niu2023empirical}, different evaluation measures~\cite{shi2022evaluation} and
interpretability~\cite{wan2022they} of BERT-related variants for Code
Summarization, and a specific research object like Copilot~\cite{mastropaolo2023robustness}. Different from these studies, our review is more comprehensive, covering a wider range of papers published over a longer period. We identified three very representative pre-trained transformer-based models for code and the top-4 most popular applications. We re-implemented all three models across all four applications and managed to contest certain beliefs regarding model architectures in the literature with their performance measured using up-to-date evaluation metrics. In addition to performance and architecture analysis, we also analyzed models' generalizability on different datasets for each application, as well as their time consumption. We additionally support our conclusions with statistical testing. Our study provides a more holistic view of the capabilities and limitations of transformer-based models in SE research. We have made all of our related code and data open-source\footnote{https://anonymous.4open.science/r/Transformer-Empirical-SE-63ED}.


In summary, our work makes the following contributions:
%
%
\begin{itemize}

\item We perform a comprehensive review of transformer-based
  research published during 2017-2022 and report on their key
  characteristics. For example, we find that the SE community lags
  other domains in adopting the latest techniques. And, we find that
  the four most popular applications of transformer-based models in SE
  are Code Summarization, Bug Fixing, Bug Detection, and Code Search,
  with 33, 29, 18, and 16 papers respectively. However, we also find that the current research overlooks some of the most critical applications that developers need.
  
\item We find that CodeBERT is best suited for understanding tasks as evidenced by the increase in evaluation metrics as well as the lowest resource consumption. Besides, we also demonstrate that CodeGPT has promising performance for specific generation tasks under the state-of-the-art metric, which refutes the claims of the incapability of decoder-only models in the literature. Also, we argue that previous studies overlook the capability of encoder-only models (CodeBERT) to complete code generation tasks, and the belief that encoder-decoder architecture is optimal for general-purpose coding tasks may not be valid.
  We also discuss how SE researchers can go about selecting models
  based on their suitability for specific tasks and some non-compliance.
  
  
\item We fill in a gap in understanding the generalizability of
  transformer-based models for SE. Our findings suggest that models trained on the benchmark and frequent dataset for Code Summarization and Bug Fixing generalize poorly to the other dataset. Nevertheless, with the experiments on mixing data from both datasets, we suggest the exploration of dataset pruning/selection techniques for future improvement.
  %
  
\item We consider the resource consumption of models and find that CodeBERT is highly efficient for understanding tasks, achieving the highest performance with the lowest resource. We question CodeT5's efficiency for generation tasks, as the highest resource consumption does not guarantee consistently better performance on different metrics.
Therefore, when
  selecting transformer-based models, researchers and practitioners
  should carefully consider the performance and time complexity
  trade-offs for their specific application.
  
\end{itemize}

%



%% file: background.tex
\section{Transformer-based Models}
\label{sec:preliminary}

The most important components of Transformer architecture are the encoder-decoder structure and attention mechanism, which resides in the Transformer blocks.

The encoder aims to extract important information from the input, and outputs the encoded representation. The encoded representation is then taken in as input by the decoder, and the decoder generates the output in an autoregressive manner~\cite{vaswani2017attention}. Some variants of the Transformer model may contain only an encoder or only a decoder.


A Transformer has multiple layers, which are called Transformer blocks, and they serve as the building blocks for the encoder and decoder. The core component of a Transformer block is the attention mechanism, which is used to process the input to each Transformer block. Through the attention mechanism, a Transformer provides context for different tokens in the input sequence.

\subsection{Pipeline for Transformer-based Pre-trained Models}

There have been many Transformer-based pre-trained models proposed over the past 5 years. They are large Transformer-based models trained on extensive amounts of unlabeled data with unsupervised objectives. The intention for developing pre-trained models is to obtain models with general and transferable knowledge in a certain domain~\cite{chakraborty2022natgen}, such as programming languages. This section presents the pipeline and variations of Transformer-based pre-trained models.

\begin{figure}[t]
 \begin{centering}
  \includegraphics[width=0.95\linewidth]{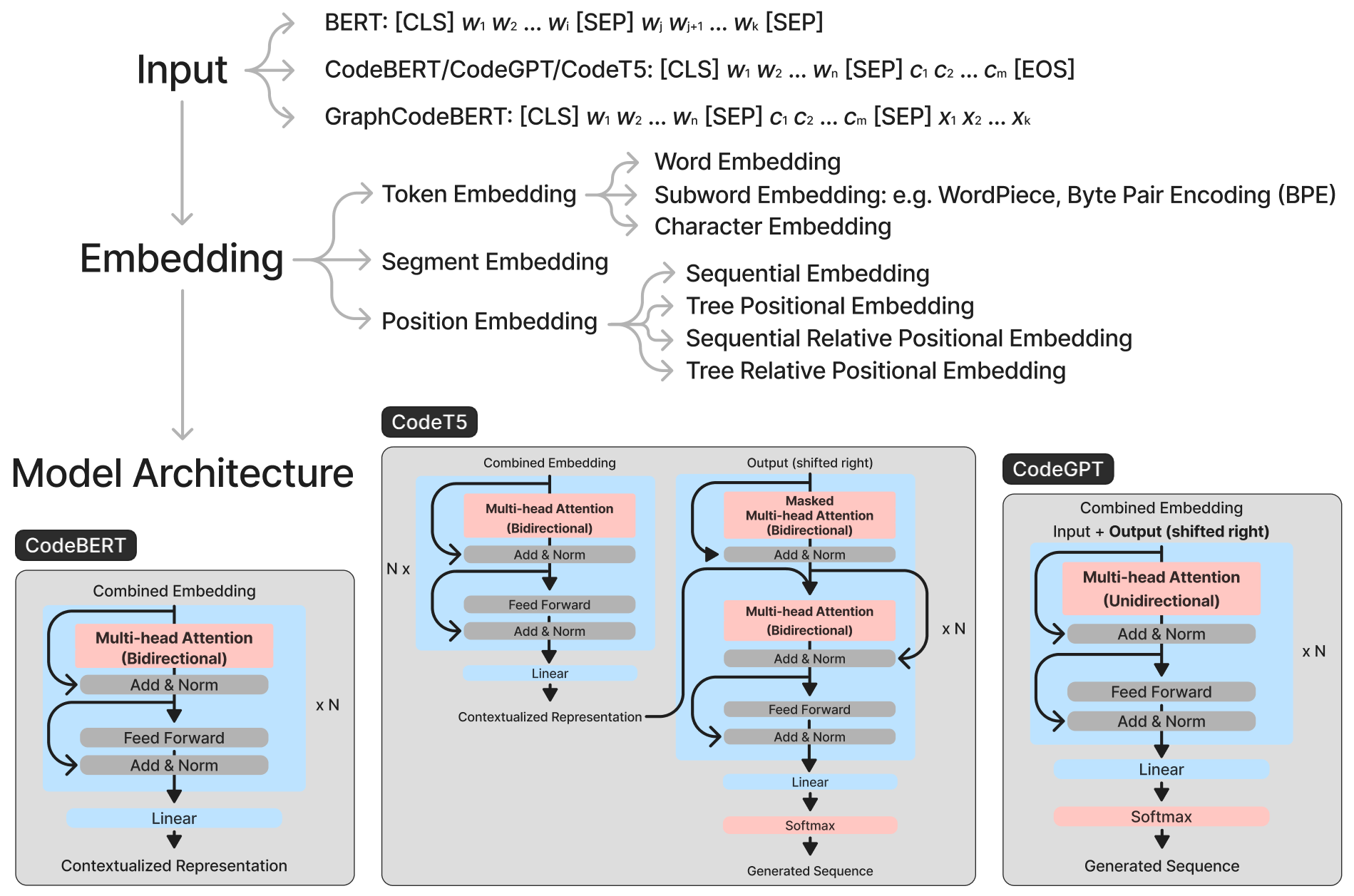}
  \par\end{centering}
 \caption{Pipeline and Variations for Transformer-based Pre-trained Models}
 \label{fig:pretrained_models}
\end{figure}

\noindent\textbf{Input:}
The inputs of Transformers vary across different models. For natural language models like BERT~\cite{kenton2019bert}, the input starts with a special token [CLS], and sentences are separated by the special separator token [SEP]. As illustrated in Figure~\ref{fig:pretrained_models}, $ w_{1} $, $ w_{2} $, …, $ w_{i} $ represents the first sentence and $ w_{j} $, $ w_{j+1} $, …, $ w_{k} $ represents the second sentence.

Language models for code (the focus of this paper), such as CodeBERT~\cite{feng2020codebert}, CodeGPT~\cite{lu2021codexglue}, CodeT5~\cite{wang2021codet5}, accept bimodal data instead, i.e., both natural language and code. $ w_{1} $, $ w_{2} $, …, $ w_{n} $ represents the natural language, and $ c_{1} $, $ c_{2} $, …, $ c_{m} $ represents the corresponding code. The two segments are separated by the [SEP] token, and [EOS] token denotes the end of input.

There exist variants of language models for code which contain more information in their input, e.g., GraphCodeBERT~\cite{guographcodebert} additionally includes variables in the input program, denoted by $ x_{1} $, $ x_{2} $, …, $ x_{k} $. Additional input combined with corresponding pre-training objectives will enhance models’ knowledge regarding a certain aspect, e.g., code structure in the case of GraphCodeBERT.

\begin{figure}[ht]
    \centering  
    \begin{lstlisting}[linewidth=6.5cm, style=mystyle]
# Calculates the area of a rectangle
def calculate_area(length, width):
    area = length * width
    return area
    \end{lstlisting}
    \caption{Original Program.}
    \label{fig:originalprogram}
\end{figure}


\newsavebox\mlm

\begin{lrbox}{\mlm}
\begin{minipage}{0.35\textwidth}
\begin{lstlisting}[style=mystyle]
# Calculates the area of a rectangle
def calculate_area(length, @[MASK]@):
    area = length * width
    return area
\end{lstlisting}
\end{minipage}
\end{lrbox}

\newsavebox\rtd

\begin{lrbox}{\rtd}
\begin{minipage}{0.35\textwidth}
\begin{lstlisting}[style=mystyle]
# Calculates the area of a rectangle
def calculate_area(length, width):
    area = length * @length@
    return area
\end{lstlisting}
\end{minipage}
\end{lrbox}

\begin{figure*}[!t]
  \centering
  \subfloat[Masked Language Modeling]{\usebox\mlm}
  \qquad
  \subfloat[Replaced Token Detection]{\usebox\rtd}
  \caption{Objectives of CodeBERT.}
  \label{fig:codebert_input}
\end{figure*}

\noindent\textbf{Input Embedding:}
Inputs with correct formats are used to generate input embeddings that are fed into the Transformer blocks. An input embedding typically consists of three parts: token embedding, segment embedding, and position embedding.
For token embedding, there exists three different options, word embedding, subword embedding, and character embedding. Among the three options, subword embedding is the most frequently used technique in Transformer-based models since it can deal with out-of-vocabulary (OOV) issues. There are two common subword tokenization methods, WordPiece~\cite{schuster2012japanese} and Byte Pair Encoding~\cite{shibata1999byte}. WordPiece is used in BERT and CodeBERT, while Byte Pair Encoding is used in CodeGPT and CodeT5.

Segment embedding is used to distinguish which segment of input a specific token belongs to. For example, in BERT, segment embedding could be 0 for words in the first sentence, and 1 for words in the second sentence.

Lastly, position embedding encodes the positional information of words. There are different options available. A traditional one is sequential embedding, for example, using \textit{sine} and \textit{cosine} functions to capture the positional information~\cite{chirkova2021empirical}. To capture structure-related positional information, tree positional embedding~\cite{shiv2019novel} can be used. Sequential relative positional embedding~\cite{shaw2018self} and tree relative positional embedding~\cite{kim2021code} captures the relative position of the input elements instead of the absolute position, which are more effective in sequence-based tasks.

\noindent\textbf{Model Architecture:}
There are three different model structures for Transformer-based models: encoder-only model, decoder-only model, and encoder-decoder model. The most representative pre-trained language models with such structures are BERT, GPT, and T5 for natural language, and CodeBERT, CodeGPT, and CodeT5 for programming languages. 

BERT~\cite{kenton2019bert}, Bidirectional Encoder Representations from Transformers, is an encoder-only model pre-trained on BookCorpus and Wikipedia, with the objective of masked language modeling and next sentence prediction~\cite{kenton2019bert}. There are multiple variants of BERT used for SE-related tasks, such as CodeBERT~\cite{feng2020codebert} and GraphCodeBERT~\cite{guographcodebert}, for type inference, automated program repair, etc.

GPT~\cite{radford2019language}, Generative Pre-trained Transformer, is a decoder-only autoregressive language model pre-trained on BookCorpus, with the generative objective of predicting the next word given some previous words~\cite{radford2019language}. GPT variants that are applied to SE-related tasks, such as code generation, include GPT-C~\cite{svyatkovskiy2020intellicode} and CodeGPT~\cite{lu2021codexglue}.

T5~\cite{raffel2020exploring} is an encoder-decoder model pre-trained on the C4 (Colossal Clean Crawled Corpus) dataset with a ``span corruption” objective~\cite{raffel2020exploring}. T5’s variants include CodeT5~\cite{wang2021codet5} and CoditT5~\cite{zhang2022coditt5} which are suitable for SE tasks such as code translation and code summarization.

CodeBERT, CodeGPT, and CodeT5 have the same model architectures as BERT, GPT, and T5 (see Figure~\ref{fig:pretrained_models}). Apart from the encoder and decoder components, the differences between the model architectures exist in the multi-head attention mechanism and input/output. For CodeBERT and CodeT5, the attention mechanism is bidirectional, allowing the model to capture the context from both directions and to better capture long-range dependencies. Whereas the attention mechanism in CodeGPT is unidirectional, which means that the model only attends to the past inputs in the sequence, avoiding potential future information leakage.

The output of CodeBERT is a contextualized representation of input, and can be utilized to perform e.g., classification tasks. For CodeT5, after the encoder has generated the contextualized representation of an input, like CodeBERT, the decoder takes it in to combine with output generated at previous steps to form the input to the decoder. CodeGPT generates output using the input combined with output generated at previous steps, and the output of CodeT5 and CodeGPT are both probabilities of tokens that can be converted to the corresponding generated sequence.

\noindent\textbf{Pre-training Objectives:}
A pre-trained model for code may have multiple objectives, which
constitute a hybrid objective function and contribute to better code
understanding~\cite{feng2020codebert,wang2021codet5}.
The three pre-trained models that we investigate in this paper have
different pre-training objectives. However, Transformer-based pre-trained models for code are mostly pre-trained on different subsets of the same dataset, CodeSearchNet~\cite{husain2019codesearchnet}, unlike the pre-trained models for natural language.

%
%

\textit{CodeBERT:} CodeBERT is pre-trained with the objectives of Masked Language Modeling (MLM) and Replaced Token Detection (RTD)~\cite{feng2020codebert}. MLM aims to predict the masked out token in both NL and PL sections of the program. Figure~\ref{fig:codebert_input}(a) is an example of MLM. The objective is to predict the original token for [MASK]. RTD aims to determine whether a token is the original one or a replaced one. For example, if the generator mutates the original program (Figure~\ref{fig:originalprogram}), to Figure~\ref{fig:codebert_input}(b), the discriminator should recognize that the ``length" is a replaced token instead of the original token.

%

\newsavebox\pnw

\begin{lrbox}{\pnw}
\begin{minipage}{0.35\textwidth}
\begin{lstlisting}[style=mystyle]
# Calculates the area of a rectangle
def calculate_area(length, width):
    area = length * @[PRED]@
\end{lstlisting}
\end{minipage}
\end{lrbox}

\begin{figure}[t]
  \centering
  \subfloat[Predicting Next Word]{\usebox\pnw}
  \qquad
  \caption{Objectives of CodeGPT.}
  \label{fig:codegpt_input}
\end{figure}

\textit{CodeGPT:} CodeGPT is pre-trained with the objective to predict the next token, given previous context. Figure~\ref{fig:codegpt_input}(a) is the illustration of the pre-training objective.

%
%
%

\newsavebox\ittf

\begin{lrbox}{\ittf}
\begin{minipage}{0.4\textwidth}
\begin{lstlisting}[style=mystyle]
 0         1      0   1  0   1  00
\end{lstlisting}
\vspace{-5pt}
\hspace{0.15cm}$\uparrow$\hspace{1.53cm}$\uparrow$\hspace{1cm}$\uparrow$\hspace{0.5cm}$\uparrow$\hspace{0.34cm}$\uparrow$\hspace{0.49cm}$\uparrow$\hspace{0.33cm}$\uparrow$$\uparrow$
\vspace{-5pt}
\begin{lstlisting}[style=mystyle]
def calculate_area(length, width):
\end{lstlisting}
\end{minipage}
\end{lrbox}

\newsavebox\mip

\begin{lrbox}{\mip}
\begin{minipage}{0.33\textwidth}
\begin{lstlisting}[style=mystyle]
def @[MASK0]@(@[MASK1]@, @[MASK2]@):
    @[MASK3]@ = @[MASK1]@ * @[MASK2]@
    return @[MASK3]@
\end{lstlisting}
\end{minipage}
\end{lrbox}

\newsavebox\bdg

\begin{lrbox}{\bdg}
\begin{minipage}{0.35\textwidth}
\begin{lstlisting}[style=mystyle]
# Calculates the area of a rectangle
\end{lstlisting}
\vspace{-6pt}
\begin{center}
$\Updownarrow$
\end{center}
\vspace{-10pt}
\begin{lstlisting}[style=mystyle]
def calculate_area(length, width):
    area = length * width
    return area
\end{lstlisting}
\end{minipage}
\end{lrbox}

\begin{figure}[!t]
  \centering
  \subfloat[Identifier Tagging]{\usebox\ittf}
  \qquad
  \subfloat[Masked Identifier Prediction]{\usebox\mip}
  \qquad
  \subfloat[Bimodal Dual Generation]{\usebox\bdg}
  \caption{Objectives of CodeT5.}
  \label{fig:codet5_input}
\end{figure}

\textit{CodeT5:} CodeT5 has four pre-training objectives. The first one is Masked Span Prediction (MSP). It can be viewed as a variation of MLM, and it allows masking of multiple consecutive tokens. Besides MSP, CodeT5 has also introduced two additional tasks: Identifier Tagging (IT) and Masked Identifier Prediction (MIP) to enable the model to learn code-specific structural information. IT aims to determine whether a token is an identifier and MIP performs obfuscation on the PL part of the program and aims to predict the masked out identifiers, as shown in Figure~\ref{fig:codet5_input}(a) and~\ref{fig:codet5_input}(b). The last pre-training objective of CodeT5 is bimodal dual generation, which aims to perform NL$\rightarrow$PL generation and PL$\rightarrow$NL generation simultaneously, as illustrated in Figure~\ref{fig:codet5_input}(c).

The different pre-training objectives, together with different model architectures, enable the models to be suitable for different tasks.

\noindent\textbf{Fine-tuning Techniques:}
Fine-tuning is a widely used technique in Transformer-based SE research. It tunes a pre-trained model, which has
been trained with unsupervised objective, on a labeled dataset to
achieve high performance on downstream tasks. Benefitting from the
general knowledge learned during the pre-training phase of model
training, fine-tuning requires a much smaller data size than
training a model from scratch.

\begin{table}
\begin{center}
	\caption{Fine-tuning and Variations}
	\label{tab:finetune}
\fontsize{9}{0}
\begin{tabular}{ c|c|c } 
 \hline
 \textbf{Tuning Methods} & \textbf{Characteristics} & \textbf{Advantage} \\ 
 \hline
 \multirow{2}{*}{Fine-tuning} & Require high-quality & Task-specific \\ 
			& labelled dataset & \& Flexible \\ 
 \hline
 \multirow{2}{*}{}Zero/Few-shot & Simulates & Better generalization \\ 
			Learning & data scarcity & ability~\cite{palatucci2009zero} \\ 
 \hline
 \multirow{2}{*}{Prompt Tuning} & Augment input & Fully utilizes pre-trained \\ 
			& with prompts & models~\cite{wang2022no,huang2022prompt} \\ 
 \hline
\end{tabular}
\end{center}
\end{table}

Despite the prevalence of fine-tuning, there exist some variants, including zero-shot, few-shot learning, and prompt tuning. Zero-shot learning applies pre-trained models without any fine-tuning and few-shot learning simulates data scarcity by providing very limited data examples. Prompt tuning transforms the downstream tasks into a similar format as the pre-training tasks using prompts. We summarize the characteristics and respective advantages of the different tuning methods in Table~\ref{tab:finetune}.


%% file: methodology.tex
\section{Methodology }
\label{sec:methodology}


Our main objective in this paper is to first provide a comprehensive
review of the transformer-based techniques proposed to tackle SE problems and published in top SE venues, we then summarize the promises and potential pitfalls of these techniques to help researchers and practitioners better understand their capabilities and limitations.

\subsection{Key Words for Literature Review}

To generate keywords for our comprehensive literature review, we first searched for the top four software engineering conferences from 2019-2022: 1) ACM Joint European Software Engineering Conference and Symposium on the Foundations of Software Engineering (ESEC/FSE) 2) International Conference on Software Engineering (ICSE) 3) International Conference on Automated Software Engineering (ASE) 4) International Symposium on Software Testing and Analysis (ISSTA). There are multiple types of pre-trained models mentioned in
the papers, including Vanilla Transformer, BERT, GPT, T5, and their variants. We recorded all the models mentioned in the papers, as well as their popular variants, resulting in a comprehensive list of 17 keywords, including Transformer, BERT, GPT, T5, CodeBERT, CodeBERTa, GraphCodeBERT, RoBERTa, CuBERT, C-BERT,
BERTOverflow, GPT-C, CodeGPT, PLBART, BART, IntelliCode, and
CodeT5.

\subsection{Identify Related Literature}

We identify 27 relevant SE conferences and journals with core ranking A* or A, as shown in Table~\ref{tab:stats}. Using the keywords mentioned above, we conducted an extensive search for Transformer-based papers in those 27 conferences and journals published between 2017 and December 2022 \footnote{We excluded ICSE'23 since the pdf of most accepted
papers were unavailable in December.}.

We locate the keywords in the papers to confirm that they are about applying Transformer-based pre-trained models to SE domain, and exclude papers that the keywords appear by coincidence: e.g., a mathematical transformer, a java library, names of authors. Note that transformer-related papers are unlikely to be missed out using the keyword list, as papers usually mention or reference the models in the list, even if the paper uses variants of them.
Finally, we identified 282 relevant papers and 57 different applications. 

\begin{table}[t]
	\begin{center}
		\caption{Statistics for Papers Published in Top-tier Venues. }
		\label{tab:stats}
		\begin{tabular}{ c|c|c|c|c|c} 
			\hline
			\textbf{Venues} & \textbf{2019} & \textbf{2020} & \textbf{2021} & \textbf{2022} & \textbf{Sum} \\ 
			\hline
			ESEC/FSE & 0 &  6 & 14 & 16 &  36 \\
			\hline
			ICSE & 0 & 4 & 15 & 21 & 40\\
			\hline
			ASE & 1 & 4 & 13 & 15 & 33 \\
			\hline
			ISSTA & 0 & 3 & 2 & 7 & 12 \\
			\hline
			TSE & 0 & 1 & 9 & 12 & 22\\
			\hline
			TOSEM & 0 & 0 & 1 & 8 & 9 \\
			\hline
			ESE & 0 & 0 & 2 & 6 & 8 \\
			\hline
			PLDI & 0 & 0 & 7 & 0 & 7\\
			\hline
			OOPSLA & 0 & 1 & 0 & 0 & 1\\
			\hline
			ISSRE & 0 & 2 & 7 & 0 & 9\\
			\hline
			ESEM & 1 & 1 & 0 & 0 & 2\\
			\hline
			SANER & 0 & 0 & 4 & 8 & 12\\
			\hline
			EASE & 0 & 0 & 1 & 1 & 2\\
			\hline
			IST & 0 & 0 & 4 & 13 & 17\\
			\hline
			JSS & 0 & 1 & 1 & 8 & 10\\
			\hline
			ICPC & 0 & 0 & 3 & 7 & 10\\
			\hline
			RE & 0 & 4 & 13 & 2 & 19\\
			\hline
			CAiSE & 0 & 1 & 2 & 1 & 4\\
			\hline
			ICSME & 1 & 4 & 11 & 0 & 16\\
			\hline
			ICST & 0 & 0 & 1 & 1 & 2\\
			\hline
			MSR & 0 & 1 & 4 & 4 & 9\\
			\hline
			ICSA & 0 & 0 & 0 & 1 & 1\\
			\hline
			ECSA & 0 & 1 & 0 & 0 & 1\\
			\hline\hline
			\textbf{Sum} & 3 & 34 & 114 & 131 & 282\\
			\hline
		\end{tabular}
	\end{center}
	\footnotesize{Note that, POPL, SEAMS, TOPLAS, and FM had 0 papers for all years.}
\end{table}

To extract information from the selected papers, we primarily focused on the pipelines used by the authors to incorporate transformers into their research. This involved studying the applications, datasets, pre-processing, input, architecture, training, and output of the transformers used. 
We also searched through all 282 papers using the 57 summarized application names and recorded the number of papers for each application. 

%% file: evaluation.tex
\section{Experiments}
\label{sec:eval}

We aim to answer the following research questions in our empirical
study.

\subsection{Research Questions}

\noindent \textbf{RQ1. Literature, Popular Applications, and Developers' Needs:} \textit{What are the characteristics of papers utilizing Transformer models, such as the yearly publication trends and the extent to which different applications have been explored?}

To summarize the publication characteristics, we have conducted a thorough review of 282 papers mentioned in Section~\ref{sec:methodology}. Based on these papers, we have analyzed and summarized all the applications
related to transformer-based models. We have examined why certain applications are more popular than others and whether they are relevant to the needs of SE developers. Through this analysis, we aim to shed light on the most important applications for SE research, and help the community focus on the areas that are most relevant to developers' needs.

\noindent \textbf{RQ2. Applications' Performance:} \textit{What is the performance of the three base models used for the top-4 popular applications?}

To investigate the performance of various models for different tasks, we conducted a comparative analysis of three representative models: CodeBERT, CodeGPT, and CodeT5. Specifically, we evaluated their effectiveness on the top-4 most popular applications: Code Summarization, Bug Fixing, Bug Detection, and Code Search. We found that previous claims about model performance and architectures have been misleading without the use of up-to-date metrics for specific tasks, and we better support our conclusion about model suitability with non-parametric Wilcoxon signed-rank test. We also identified the most commonly-used models for each task by reviewing the literature, and checked for consistency with our findings.

\noindent \textbf{RQ3. Generalization:} \textit{How well do the base models trained on commonly-used benchmark datasets for each application generalize to frequent datasets, and conversely, how well do models trained on frequent datasets perform on the benchmark datasets?
}

To evaluate the generalization ability of the models, we conducted a search for the datasets used in all papers related to the top-4 applications. For these applications, we adopt the dataset from CodeXGLUE~\cite{lu2021codexglue} as the benchmark dataset, as CodeXGLUE is widely used by Transformer-based techniques for SE applications. The frequent dataset is defined as the dataset with the highest \emph{frequency} of use other than the benchmark dataset. We follow the common train test split used in the literature. We found that the benchmark datasets were the most commonly used in all applications except for Bug Fixing, where more researchers utilized datasets such as Defects4J~\cite{just2014defects4j} than the benchmark dataset provided (BFPsmall \& BFPmedium~\cite{tufano2019empirical}). However, for Code Search and Bug Detection, only a few researchers used non-benchmark datasets, which makes it less meaningful to investigate generalization on those datasets. Therefore, we focused on the benchmark and frequent datasets for Code Summarization and Bug Fixing. We then evaluate the models trained on the benchmark dataset on the test set of the frequent dataset and vice versa to examine their generalization capabilities. Statistical testing is performed to safeguard our conclusions.

\noindent \textbf{RQ4. Resource Consumption:} \textit{What is the resource consumption of inference for each base model and application?}

We answer this question by comparing and analyzing the average inference time and memory usage of various models on all 4 applications across different datasets.

\subsection{Experimental Setup}

\noindent\textbf{Pre-trained Models.} We choose CodeBERT, CodeGPT, and CodeT5 as the pre-trained models with the consideration of representativeness in model architectures and their wide presence in the literature.

\textit{CodeBERT} is an encoder-only model which has the same model architecture as RoBERTa~\cite{liu2019roberta}. It is pre-trained on CodeSearchNet and is capable of processing both source code and natural language text. The model we use is CodeBERT-base, which has 125M parameters\footnote{https://huggingface.co/microsoft/codebert-base}.

\textit{CodeGPT} is a decoder-only model which has the same model architecture as GPT-2~\cite{radford2019language}. The model we use in our paper is CodeGPT-small-java-adaptedGPT2\footnote{https://huggingface.co/microsoft/CodeGPT-small-java-adaptedGPT2} with 124M parameters, which is pre-trained on the Java corpora from the CodeSearchNet dataset and uses the GPT-2 model as the starting point.

\textit{CodeT5} is a variant of T5~\cite{raffel2020exploring}, and achieves state-of-the-art performance for many code intelligence tasks. It views all tasks through a sequence-to-sequence paradigm, and can solve both code understanding and generation tasks. CodeT5 is pre-trained on CodeSearchNet and an additional C/C\# dataset~\cite{wang2021codet5}. We use CodeT5-base\footnote{https://huggingface.co/Salesforce/codet5-base} that has 220M parameters.


\vspace{.2em}
\noindent\textbf{Datasets.} Table~\ref{tab:dataset} shows the benchmark datasets we use in the experiments for the four applications.

\begin{table}[t]
	\caption{Datasets.}		
	\label{tab:dataset}	
	\centering{}%
	\resizebox{\linewidth}{!}{\begin{tabular}{c|c|c}
		\hline
		\textbf{Task} & \textbf{Fine-tuning Dataset} & \textbf{Train/Valid/Test} \tabularnewline 
		\hline
		Code Summarization & Java Subset in CodeSearchNet~\cite{husain2019codesearchnet} & 164,923 / 5,183 / 10,955 \tabularnewline 
		\hline
		\multirow{2}{*}{Bug Fixing} & BFPsmall~\cite{tufano2019empirical} & 46,680 / 5,835 / 5,835 \tabularnewline
        \cline{2-3}
		& BFPmedium~\cite{tufano2019empirical} & 52,364 / 6,545 / 6,545 \tabularnewline
		\hline
		Bug Detection & Zhou et al.~\cite{zhou2019devign} & 21,854 / 2,732 / 2,732 \tabularnewline
		\hline
		Code Search & Python Subset in CodeSearchNet~\cite{husain2019codesearchnet} & 251,820 / 9,604 / 19,210  \tabularnewline
		\hline
	\end{tabular}}
\end{table}

\textit{Code Summarization:} CodeSearchNet~\cite{husain2019codesearchnet} is a dataset which consists of $<code, comment>$ pairs from open source projects. Code refers to the code snippet for a method, and comment refers to the description of the code, for example in Javadoc format.

\textit{Bug Fixing:} This dataset is provided by Tufano et al.~\cite{tufano2019empirical}. It contains method-level pairs of the buggy and corresponding fixed code from thousands of GitHub Java repositories. The pairs are called bug-fix pairs, BFPs in short. Based on the code length, Tufano et al. provided two datasets: BFPsmall, which has a code length below 50; and BFPmedium, which has a length between 50 and 100.

\textit{Bug Detection:} This dataset is provided by Zhou et al.~\cite{zhou2019devign}. It contains 27k+ C code snippets from two open-source projects, FFmpeg and QEMU, of which 45\% are defective.

\textit{Code Search:} This dataset is the Python version of the CodeSearchNet~\cite{husain2019codesearchnet} dataset.

\vspace{.2em}
\noindent\textbf{Evaluation Metrics.}
Different metrics are used for the four applications studied in this paper. 

\textit{Code Summarization:} Following previous works~\cite{feng2020codebert,shi2022we}, we use BLEU~\cite{papineni2002bleu}, METEOR~\cite{banerjee2005meteor}, and ROUGE-L~\cite{lin2004rouge} to measure the quality of the summary generated in the Code Summarization task. Each of them considers different aspects.

BLEU is based on the n-gram precision between the generated summary and the reference~\cite{papineni2002bleu}: 
\begin{equation}
	BLEU = BP*exp(\sum^{N}_{n=1}w_{n}logp_{n})
	\label{eq:bleu}
\end{equation}
where \textit{BP} is the brevity penalty that penalizes short summary. $p_{n}$ refers to n-gram precision and $w_{n}$ is the weight.

METEOR focuses on the harmonic mean of unigram precision and recall\cite{banerjee2005meteor}:
\begin{equation}
	METEOR = (1-P)*F_{mean}
	\label{eq:meteor}
\end{equation}
where \textit{P} is the penalty for difference between the word order in the summary generated and the reference, and more weight is put on recall when calculating the harmonic mean $F_{mean}$. METEOR allows exact, stem, and synonym matches.

ROUGE-L is based on the longest common sub-sequence (LCS) between the generated summary and the reference~\cite{lin2004rouge}:
\begin{equation}
	ROUGE-L = \frac{(1+\beta^2)R_{lcs}P_{lcs}}{R_{lcs}+\beta^2P_{lcs}}
	\label{eq:rouge}
\end{equation}
where $R_{lcs}$ measures the proportion of the LCS length relative to the length of the reference and $P_{lcs}$ measures that to the length of the generated summary, and ROUGE-L calculates the harmonic mean of them.

\textit{Bug Fixing:} We use BLEU-4, Accuracy, and CodeBLEU to measure the quality of the repaired code, where accuracy considers only exact matches, and CodeBLEU~\cite{ren2020codebleu} additionally considers aspects such as code structure:
\begin{equation}
 \begin{split}
	CodeBLEU &= \alpha*BLEU + \beta*BLEU_{weight} \\
             & + \gamma*Match_{ast} + \delta*Match_{df}
	\label{eq:codebleu}
 \end{split}
\end{equation}
which involves n-gram, weighted n-gram, AST, and data-flow matches.

The definition of accuracy adopted in this paper is the same as below except that $y_{i}$ and $\hat{y}_{i}$ refer to the buggy and fixed code instead.

\textit{Bug Detection:} We use Accuracy as the evaluation metric for bug detection, following the work of CodeT5~\cite{wang2021codet5}. Accuracy helps to measure the ability of the model to distinguish buggy code from normal code~\cite{wang2022no},
\begin{equation}
	Accuracy = \frac{\sum_{i=1}^{|D|}1(y_{i}==\hat{y}_{i})}{|D|}
	\label{eq:acc}
\end{equation}
where \textit{D} refers to the dataset, and $|D|$ refers to its size, $y_{i}$ and $\hat{y}_{i}$ refer to the ground truth label and predicted label, respectively. The function in the numerator is \textit{1} if the two labels are equal, and 0 otherwise.

\textit{Code Search:} We use Mean Reciprocal Rank (MRR) to measure the ability of the model to retrieve relevant code given a natural language query. It calculates the multiplicative inverse of the rank of the correctly retrieved code snippet and is defined as below~\cite{hu2022lighting}.
\begin{equation}
	MRR = \frac{1}{|Q|}\sum_{i=1}^{|Q|}\frac{1}{rank_{i}}
	\label{eq:mrr}
\end{equation}
where $rank_{i}$ refers to the rank of the first correctly retrieved code snippet and $|Q|$ represents the number of queries.

\vspace{.2em}
\noindent\textbf{Configurations.}
We conducted all experiments on a NVIDIA RTX A4000 GPU with CUDA version 11.6 and 16GB of VRAM. Implementation of all models are under the PyTorch framework.

\section{Results }
\label{sec:results}

In this section, we aim to answer the research questions mentioned in Section~\ref{sec:eval} by discussing the promises and perils of using transformer-based models for SE research.

\subsection{RQ1. Literature, Popular Applications, and Developers' Needs}

\noindent\textbf{On Papers Published.}
Table \ref{tab:stats} presents the number of papers on Transformer-based techniques published in top-tier SE conferences or journals during 2019-2022. 
Through our search described in Section~\ref{sec:methodology}, we found a total of 3 papers published in 2019, 34 papers in 2020, 114 papers in 2021, and 131 papers in 2022. The increasing trend in the numbers indicates a growing interest in using Transformer-based techniques to solve SE-related problems.

\begin{figure}[t]
	\centering
        \includegraphics[width=7.6cm]{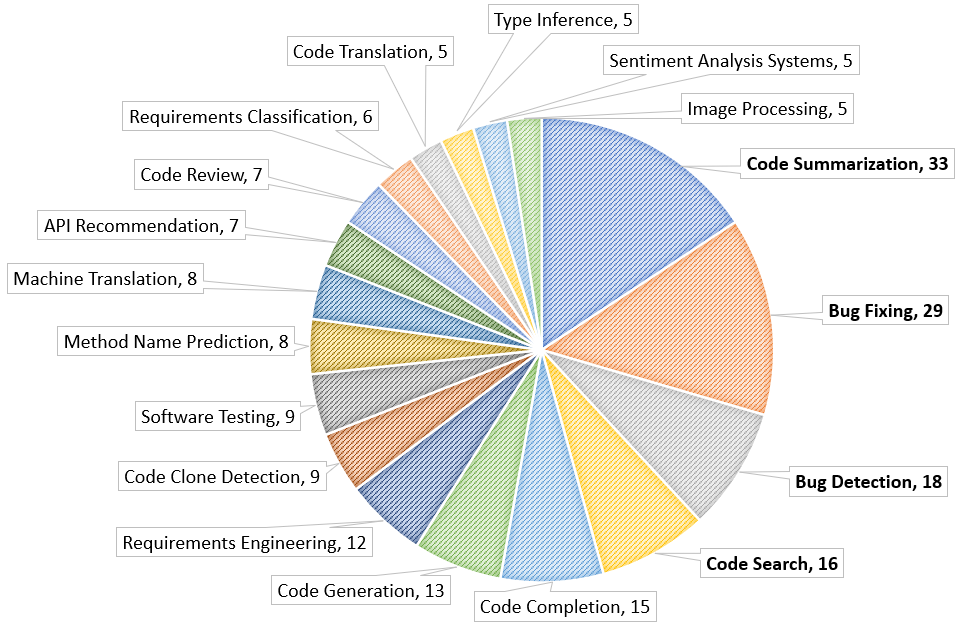} 
	\caption{Frequency of Applications. We only list applications with frequency~$\ge$~5.}
	\label{fig:app}
\end{figure}

Although the Transformer architecture was proposed in 2017~\cite{vaswani2017attention}, it was rarely used by SE researchers until 2020. Besides, even though transformer models are more widely studied over time, there is still room for improvement. For example, some SE studies~\cite{gao2022gt,zhu2022simple} only compare their models to Vanilla Transformers instead of including additional and more advanced models such as T5, nearly three years after the proposal of these new models. This shows that the SE research community needs to explore breakthrough techniques from other domains and investigate their application to SE-related tasks. Additionally, researchers should stay informed and attentive to the latest methodologies, such as T5, CodeT5 and so on. 

\begin{tcolorbox}[colframe=gray,left=6pt,right=6pt,boxrule=1.5pt,arc=.5em]\textit{\textbf{Peril 1:} SE researchers using transformer-based models should pay attention to recent techniques in domains like NLP and be more attentive to the latest methodologies.}
\end{tcolorbox}

\noindent\textbf{On Applications.}

We collected an exhaustive list of applications explored by Transformer-related papers in ICSE/FSE/ASE/ISSTA by 2022. We further examined the frequency of the applications in papers in Table \ref{tab:stats}. Parts of the application statistics are shown in Figure~\ref{fig:app}. Out of a total of 282 papers, we found that code summarization appeared 33 times, bug fixing appeared 29 times, bug detection appeared 18 times, and code search appeared 16 times, making them the top four popular applications. Due to their representation of the community's research focus, we chose to conduct experiments on these four applications for our research questions.

\begin{table}
	\begin{center}
		\caption{Top-4 Popular Applications}
		\label{tab:top4app}
		\resizebox{\linewidth}{!}{\begin{tabular}{ c|c|c } 
				\hline
				\textbf{Task Name} & \textbf{Nature} & \textbf{Category} \\ 
				\hline
				Code Summarization & Code $\rightarrow$ NL & Generation \\
				\hline
				Bug Fixing & Code $\rightarrow$ Code & Generation \\
				\hline
				Bug Detection & Code $\rightarrow$ Class (0/1) & Understanding \\
				\hline
				Code Search & NL $\rightarrow$ Code & Understanding \\
				\hline
		\end{tabular}}
	\end{center}
\end{table}

\textit{Code Summarization}~\cite{iyer2016summarizing}: This task generates a natural language summary for a code snippet to aid programmers' understanding.

\textit{Bug Fixing}~\cite{tufano2019empirical}: This task aims to fix buggy programs automatically.

\textit{Bug Detection}~\cite{zhou2019devign}: This task performs binary classification - determining if a program is buggy or not.

\textit{Code Search}~\cite{husain2019codesearchnet}: This task searches for relevant code snippets given their natural language descriptions.

The four tasks belong to different categories (Table~\ref{tab:top4app}), making them representative of SE tasks researchers study and appropriate for our empirical study. Furthermore, due to their specific task implementation, bug detection and code search tend to be more focused on understanding tasks, while code summarization and bug fixing are more generation-oriented. Thus, the optimal performance on the tasks is achieved by different types of models, as discussed in RQ2.

\begin{tcolorbox}[colframe=gray,left=6pt,right=6pt,boxrule=1.5pt,arc=.5em]\textit{\textbf{Promise 1:} The four most popular tasks, in descending order of popularity, are Code Summarization, Bug Fixing, Bug Detection, and Code Search. The first two are generation tasks and the next two are understanding tasks; all of these tasks benefit from transformer-based techniques.}
\end{tcolorbox}

Apart from experimenting with the most representative and popular applications, we have also identified the least-studied applications. Some of the applications are too specific, e.g., taint propagation detection and defect inheritance reduction, leading to fewer researchers studying them. Some are related to the more popular applications, e.g., algorithm classification can be seen as the derivative of code summarization, as the core idea of both applications is to understand intent. One potential reason for other applications to be less well-studied is that there are no commonly used benchmarks.

\begin{tcolorbox}[colframe=gray,left=6pt,right=6pt,boxrule=1.5pt,arc=.5em]\textit{\textbf{Peril 2:} The least-studied applications have the following characteristics: too specific, related to more popular applications, lacking high quality benchmarks.}
\end{tcolorbox}


\noindent\textbf{Developers' Needs.} Existing studies have emphasized the importance of communication and collaboration in the software development process~\cite{gonccalves2011collaboration,aniche2018modern}. However, the potential of transformer-based techniques, such as ChatGPT, to facilitate these processes is yet to be extensively explored. Additionally, the time spent by developers in seeking information is a critical factor in their daily work~\cite{gonccalves2011collaboration,ko2007information}. While code search has been identified as one of the top-4 most popular applications, there are still many detailed issues that developers frequently encounter that remain unexplored. For example, how to efficiently transfer deprecated features, functions, or methods to new ones~\cite{sawant2018understanding}, or how to solve module dependencies~\cite{bogart2015breaks}. Furthermore, despite code summarization being the top-1 most popular application, it is rated much lower than commit conflict resolution by developers~\cite{treude2015summarizing}. Surprisingly, the frequency of transformer-based techniques being used for commit-related applications is much lower in our study (i.e., code summarization: 33, conflict resolution: 2).

\begin{tcolorbox}[colframe=gray,left=6pt,right=6pt,boxrule=1.5pt,arc=.5em]\textit{\textbf{Peril 3:} Current research neglects applications actually demanded by developers. SE researchers should prioritize applications that are most relevant to developers' needs, such as improving communication and collaboration, efficiently transferring deprecated features/functions/methods to new ones, solving module dependencies and dealing with commit conflicts.}
\end{tcolorbox}

\subsection{RQ2. Applications' Performance}

\noindent\textbf{Suitability for Different Tasks.} The performance of the three representative models - CodeBERT, CodeGPT, and CodeT5, on the four different tasks - Code Summarization, Bug Fixing, Bug Detection, and Code Search is summarized in Table~\ref{tab:application}. Our implementation has comparable performance to the literature~\cite{lu2021codexglue,zeng2022extensive,niu2022spt,chakraborty2022natgen}. We can observe from Table~\ref{tab:application} that the highest performance on each of the four applications is achieved by different models. 

\begin{table}[t]
	\caption{Performance of three representative models on top-4 applications.}		
	\label{tab:application}	
	\centering{}%
	\resizebox{\linewidth}{!}{\begin{tabular}{c|c|c|c|c|c}
		\hline
		& \textbf{Code Summarization}  & \multicolumn{2}{c|}{\textbf{Bug Fixing}} &  \textbf{Bug} &  \textbf{Code} \tabularnewline
		& \textbf{(BLEU-4/} & \multicolumn{2}{c|}{\textbf{(BLEU-4/Accuracy/CodeBLEU)}} &  \textbf{Detection} &  \textbf{Search} \tabularnewline
				\cline{3-4}
		& \textbf{METEOR/ROUGE-L)} & Small (S) & Medium (M) &  \textbf{(Accuracy)} &  \textbf{(MRR)} \tabularnewline
		\hline 
		CodeBERT & 18.74/13.16/35.03 & 74.41/17.58/\textbf{79.04} & 88.60/10.10/\textbf{88.40} & \textbf{63.54} & \textbf{28.45} \tabularnewline
		\hline 
		CodeGPT & 14.15/\textbf{15.89}/33.23 & 72.84/19.78/77.31 & 86.56/12.07/86.09 & 63.25 & 23.93 \tabularnewline
		\hline 
		CodeT5 & \textbf{20.20}/15.06/\textbf{38.20} & \textbf{78.04/21.54}/77.70 & \textbf{88.86/13.60}/86.42 & 62.99 & 27.41 \tabularnewline
		\hline
	\end{tabular}}
\end{table}

For Bug Detection and Code Search, CodeBERT achieves the best performance, whereas, the best performance for Code Summarization and Bug Fixing is achieved by different models under different metrics.

Bug Detection is an understanding task, as once the model understands the code, it will be able to predict whether the code is defective or not. Our implementation of Code Search is to compare the vector representations of code candidates and the natural language query, which makes the task fall into the category of understanding as well. With a p-value of 0.05, CodeBERT significantly outperforms CodeGPT and CodeT5 in Code Search. Although the advantage in Bug Detection is not statistically significant, CodeBERT still remain as the best choice with lowest resource consumption, as discussed in RQ4. Thus, consistent with the literature\cite{zeng2022extensive}, we conclude that an encoder-only model like CodeBERT is suitable for understanding tasks.

For Code Summarization, CodeT5 achieves the highest BLEU-4 and ROUGE-L, while the higher METEOR achieved by CodeGPT is also statistically significant. METEOR is a metric that better correlates with human judgment\cite{banerjee2005meteor}, as it considers aspects such as synonyms during evaluation. Previous experiments~\cite{roy2021reassessing} have also demonstrated that METEOR is a more reliable metric than commonly-used metrics such as BLEU-4, and it has become the state-of-the-art evaluation metric for Code Summarization tasks. Thus, this result allows us to refute the claim~\cite{zeng2022extensive} that decoder-only models (such as CodeGPT) fail to enable optimal performance on any task, as the claim relies only on outdated metric, BLEU-4, to draw the conclusion. 

For Bug Fixing, CodeT5 achieves the highest BLEU-4 and Accuracy, whereas CodeBERT achieves the highest CodeBLEU. CodeBLEU is a metric that takes aspects such as code structure into consideration and is arguably more important than BLEU-4 and Accuracy, which focus on term matching in the Bug Fixing task. Moreover, CodeT5's better performance on BLEU-4 on the BFPmedium dataset is insignificant, whereas CodeBERT significantly outperforms CodeT5 in terms of CodeBLEU. Thus, we conclude that encoder-only models are able to compete with and even outperform encoder-decoder models on specific program generation tasks, while achieving the highest performance in code-understanding tasks. This has been overlooked by previous studies that believe the encoder-decoder architecture is the most suitable architecture to build general-purpose pre-trained models that perform both understanding and generation tasks~\cite{ahmad2021unified,lewis2019bart,raffel2020exploring}. Also, the higher performance achieved by CodeT5 in BLEU-4 and Accuracy may not necessarily depend on the model architecture, as the pre-training objectives of CodeT5 incorporate specifically designed code-related tasks, which may have helped it learn term matching and perform better. 

Thus, future work can be done to verify if encoder-only models can perform better at code generation tasks if more sophisticated code-related and generative pre-training objectives are used.



\noindent\textbf{Most Frequently Used Pre-trained Model for Each Task.} We extracted the information from every paper with any one of the four applications. Table~\ref{tab:app_sum} is the summary of the most frequently used pre-trained model, as well as the best-performing model for each task.

\begin{table}
	\begin{center}
		\caption{Most Frequently Used Model vs. Best Performing Model.}
		\label{tab:app_sum}
		\begin{tabular}{ c|c|c } 
			\hline
			\multirow{2}{*}{\textbf{Task Name}} & \textbf{Most Frequent} & \textbf{Best Performing} \\ 
			& \textbf{Model} & \textbf{Model} \\ 
			\hline
			Code Summ- & Vanilla Trans- & \multirow{2}{*}{CodeGPT/CodeT5} \\
			arization & former (18/33) & \\
			\hline
			Bug Fixing & CodeBERT (13/29) & CodeBERT/CodeT5 \\
			\hline
			Bug Detection & CodeBERT (8/18) & CodeBERT \\
			\hline
			Code Search & CodeBERT (7/16) & CodeBERT \\
			\hline
		\end{tabular}
	\end{center}
\end{table}
We can see that the most frequently used model is also the best-performing model for Bug Detection and Code Search, and partially for Bug Fixing. However, this is not the case for Code Summarization. 

For Code Summarization, CodeT5 is a more advanced pre-trained model with an encoder-decoder architecture, just like Vanilla Transformer. CodeGPT is a decoder-only model with a different model architecture. Previous studies~\cite{lu2021codexglue,wang2021codet5} collectively demonstrate that CodeT5 is able to outperform the Vanilla Transformer significantly in this task, and our result shows that CodeGPT has competitive performance over CodeT5 under state-of-the-art evaluation metric.
Thus, the community should watch for ML and SE advancements and integrate advanced models to achieve optimal results, instead of using the earliest or maybe the most well-known models.

For Bug Fixing, it is worth noting that the literature commonly believes that the encoder-decoder models are more suitable for code generation tasks~\cite{ahmad2021unified,lewis2019bart,raffel2020exploring}, yet many papers have opted for encoder-only models (CodeBERT) without proper justification or comparison to encoder-decoder models.

\begin{tcolorbox}[colframe=gray,left=6pt,right=6pt,boxrule=1.5pt,arc=.5em]\textit{\textbf{Peril 4:} Encoder-only model (such as CodeBERT) is more suitable for understanding tasks, and decoder-only model (such as CodeGPT) and encoder-only model (such as CodeBERT) can also outperform encoder-decoder model (such as CodeT5) under different metrics for generation tasks. Previous claims regarding the incapability of decoder-only models and the optimality of encoder-decoder models do not hold. Also, the most frequently used model for generation tasks is Vanilla Transformer (18 out of 33), indicating that the community should put care into selecting the most suitable models on a task-by-task basis.}
\end{tcolorbox}

\subsection{RQ3. Generalization}

In this research question, we explore how well the base models trained on commonly-used benchmark datasets for each application generalize to frequent datasets, and conversely, how well models trained on frequent datasets perform on the benchmark datasets. Table~\ref{tab:alt_data} presents the benchmark and frequent datasets used in our study. For Code Summarization, the frequent dataset was used 6 times (compared to 11 times for the benchmark dataset), while for Bug Fixing, the frequent dataset was used 5 times (compared to 4 times for the benchmark dataset). To ensure the validity of comparison, we pre-processed the frequent datasets to be in the same format as benchmark datasets. For example, patches in Defects4J were processed into bug-fix pairs.

\begin{table}
	\begin{center}
		\caption{Frequency of Datasets.}
		\label{tab:alt_data}
		\resizebox{\linewidth}{!}{\begin{tabular}{ c|c|c } 
			\hline
			\textbf{Task Name} & \textbf{Benchmark Dataset} & \textbf{Frequent Dataset} \\ 
			\hline
			Code Summarization & CodeSearchNet (11) & LeClair et al.~\cite{leclair2019recommendations} (6) \\
			\hline
			Bug Fixing & BFPmedium (4) & Defects4J, etc\footnote{The frequent datasets for Bug Fixing include Defects4J, Bugs.jar, Bears, QuixBugs, and ManySStuBs4J.} (5). \\
			\hline
		\end{tabular}}
	\end{center}
        \footnotesize{Code Search and Bug Detection are less meaningful as only a few researchers used non-benchmark datasets.}
\end{table}

To assess the generalization ability of the models trained on the benchmark and frequent datasets, we conducted experiments where we trained each model on the benchmark dataset and then evaluated it on the test set of the frequent dataset. This enabled us to evaluate whether the knowledge that the models learned from the benchmark dataset is transferable to other datasets or if it is only useful for the benchmark dataset. We also evaluated the performance of the model trained on the frequent dataset on the test set of the benchmark dataset and compared it to the performance of the model trained on the benchmark dataset itself.

\noindent\textbf{Code Summarization.} Table~\ref{tab:gen_summ} presents the performance of various models on different datasets and settings. Note that ``B” refers to the benchmark dataset, ``F” refers to the frequent dataset, ``Mix-Bench” consists of the benchmark dataset combined with randomly sampled data from frequent dataset (the sampled dataset has roughly equal size as the benchmark dataset). ``B model on B” refers to the performance of model trained on benchmark dataset evaluated on the corresponding testing set. ``F model on B” refers to the performance of model trained on the frequent dataset evaluated on the testing set of benchmark dataset, etc.

\begin{table*}[t]
	\caption{Generalization Performance of Models on Code Summarization (BLUE-4/METEOR/ROUGE-L).}		
	\label{tab:gen_summ}	
	\centering{}%
	\resizebox{\linewidth}{!}{\begin{tabular}{c|c|c|c|c|c}
		\hline
		\textbf{Code} & \textbf{B model} & \textbf{F model} & \textbf{Mix-Bench} & \textbf{F model} &  \textbf{B model} \tabularnewline
		  \textbf{Summarization} & \textbf{on B} & \textbf{on B} & \textbf{model on B} & \textbf{on F} & \textbf{on F} \tabularnewline
		\hline 
		CodeBERT & 18.74/13.16/35.03 & 8.70/5.29/14.46 & 19.31/13.25/35.73 & 32.20/20.52/42.17 & 17.61/12.87/25.75 \tabularnewline
		\hline 
		CodeGPT & 14.15/15.89/33.23 & 2.70/1.91/7.81 & 13.92/15.41/32.80 & 31.97/20.19/41.83 & 12.85/13.19/22.89 \tabularnewline
		\hline 
		CodeT5 & 20.20/15.06/38.2 & 12.10/8.92/21.53 & 20.52/14.51/38.08 & 32.94/22.05/44.33 & 19.24/14.25/28.35 \tabularnewline
		\hline
	\end{tabular}}
\end{table*}

When evaluating the models trained on either the benchmark or the frequent dataset on the test set of the other dataset, we found that they both do not generalize well (refer to columns ``B model on B", ``F model on B", ``F model on F", and ``B model on F" in Table~\ref{tab:gen_summ}). For instance, when evaluating CodeBERT trained on the benchmark dataset on the frequent dataset, its BLEU-4 score is only 17.61, which is significantly lower than the BLEU-4 score of 32.20 achieved by the model trained on the frequent dataset itself. Across all models and evaluation metrics, B model on B is able to significantly outperform F model on B, and F model on F significantly outperforms B model on F. Thus, both benchmark and frequent dataset fail to enable models to generalize onto other datasets.

Therefore, we investigated if combining data from both datasets could enable the model to learn more diverse knowledge. 

The performance of the models trained on the Mix-Bench dataset and evaluated on the benchmark dataset is shown in the ``Mix-Bench model on B" column of Table~\ref{tab:gen_summ}. Through statistical testing, we found only CodeBERT's BLEU-4 and ROUGE-L have been significantly improved. The improvement in CodeBERT's METEOR and CodeT5's BLEU-4 are insignificant. Moreover, additional training data has led to significant performance drop in CodeGPT across all metrics, and CodeT5 on METEOR. Therefore, we can conclude that including more diverse data into the training dataset does not necessarily increase model generalization ability. We thus suggest future work look into the possibility of utilizing dataset pruning/selection techniques to improve the dataset quality, which in turn will increase model generalization ability.

We observed that adding information from the frequent dataset improved the performance of CodeBERT on the benchmark dataset. However, since the mixed dataset is twice as large as the benchmark dataset, the comparison may not be fair. To address this, we created the ``Mix-Half” dataset, which has approximately the same size as the benchmark dataset. Its performance is presented in the last two columns of Table~\ref{tab:gen_summ}.



\noindent\textbf{Bug Fixing.} Table~\ref{tab:gen_bug} presents the results of evaluating models trained on Benchmark/Frequent datasets on the Frequent/Benchmark test sets for Bug Fixing. With statistical testing, we found that benchmark model's performance on benchmark dataset is significantly higher than frequent model's performance on benchmark dataset, across all three models and different metrics. The same holds true for frequent model's performance on frequent dataset and benchmark model's performance on frequent dataset. Thus, we conclude that the benchmark and frequent dataset of Bug Fixing also fail to enable model generalization ability onto other datasets.

\begin{table}[t]
	\caption{Generalization Performance of Models on Bug Fixing (BLEU-4/Accuracy/CodeBLEU).}		
	\label{tab:gen_bug}	
	\centering{}%
	\resizebox{\linewidth}{!}{\begin{tabular}{c|c|c|c|c}
		\hline
		\textbf{Bug Fixing} & \textbf{Benchmark (M)} & \textbf{Frequent} & \textbf{B model on F} & \textbf{F model on B} \tabularnewline
		\hline 
		CodeBERT & 88.60/10.10/88.40 & 97.61/92.87/92.73 & 13.29/0/18.96 & 88.04/0.05/81.45 \tabularnewline
		\hline 
		CodeGPT & 86.56/12.07/86.09 & 97.16/92.20/92.24 & 14.55/0/16.99 & 87.41/0.08/84.91 \tabularnewline
		\hline 
		CodeT5 & 88.86/13.60/86.42 & 90.04/90.37/92.21 & 8.16/0/32.84 & 67.10/0.00/76.35 \tabularnewline
		\hline
	\end{tabular}}

\footnotesize{\vspace{0.5em}Note: the zeroes in the table indicate that there are no exact matches, however, partial matches exist and are reflected by the BLEU-4/CodeBLEU scores.}
\end{table}

\begin{tcolorbox}[colframe=gray,left=6pt,right=6pt,boxrule=1.5pt,arc=.5em]\textit{\textbf{Peril 5:} For Code Summarization and Bug Fixing, both benchmark and frequent dataset fail to enable models to generalize onto other datasets. Furthermore, we found that including more data does not necessarily improve model generalization ability, and suggest the exploration of dataset pruning/selection techniques for future improvement.}
\end{tcolorbox}

%

\subsection{RQ4. Resource Consumption}

In this research question, we investigate the resource consumption of different models across tasks and datasets. Resource consumption becomes important if a model is deployed and concurrently used by many users. We focus on resource consumption during the inference phase for this reason. Table~\ref{tab:time} presents the average inference time and memory consumption for the three models across tasks and datasets. 
 
\begin{table}[t]
	\caption{Inference Time and Memory Consumption.}		
	\label{tab:time}	
	\centering{}%
	\resizebox{\linewidth}{!}{\begin{tabular}{c|c|c|c|c|c|c|c|c}
		\hline
		\textbf{Time (seconds)} &  \multicolumn{3}{c|}{\textbf{Code Summarization}}  & \multicolumn{3}{c|}{\textbf{Bug Fixing}} & \textbf{Bug} & \textbf{Code}\tabularnewline
		\cline{2-7}
		\textbf{Memory (GB)} & Benchmark  & Frequent & Mix-Bench & Benchmark (S)$^a$ & Benchmark (M)$^b$ & Frequent & \textbf{Detection} & \textbf{Search} \tabularnewline
		\hline 
		CodeBERT & 0.27/0.72 & 0.20/0.72 & 0.22/0.72 & 0.55/0.71 & 1.55/0.71 & 0.78/0.71 & 0.13/0.51 & 0.14/0.51 \tabularnewline
		\hline 
		CodeGPT & 0.34/0.52 & 0.19/0.52 & 0.26/0.52 & 0.55/0.51 & 1.07/0.51 & 0.68/0.51 & 0.14/0.52 & 0.14/0.52 \tabularnewline
		\hline 
		CodeT5 & 0.26/0.90 & 0.25/0.90 & 0.26/0.90 & 0.87/0.89 & 1.70/0.89 & 1.00/0.89 & 0.17/0.90 & 0.15/0.90 \tabularnewline
		\hline
	\end{tabular}}
\footnotesize{\vspace{0.5em}$^a$ and $^b$ stand for BFPsmall and BFPmedium.}   
\end{table}

We conclude that CodeBERT is highly efficient for understanding tasks like Bug Detection and Code Search, achieving the highest performance while consuming the least resources. Additionally, significantly more resources are required for CodeBERT to perform generation tasks than understanding tasks.

For generation tasks like Code Summarization and Bug Fixing, while CodeT5 demonstrates superior performance in certain metrics, it also exhibits an increase in resource consumption attributable to the model's complexity. This raises questions regarding the efficiency of CodeT5 in generation tasks, especially given its higher resource consumption and subpar performance observed in some experiments employing more targeted and contemporary metrics, such as CodeBLEU and METEOR. Additionally, CodeT5 consistently utilizes more memory compared to CodeBERT and CodeGPT across all tasks, a consequence of its larger parameter size, standing at 220M.

Besides, with the contrast between the time consumption for Benchmark (S) and Benchmark (M) for Bug Fixing, which differ in the length of code, we can clearly see that the time complexity of models for a certain task increases when the input complexity increases.


\begin{tcolorbox}[colframe=gray,left=6pt,right=6pt,boxrule=1.5pt,arc=.5em]\textit{\textbf{Promise 2:} CodeBERT is the most efficient model for code understanding tasks, achieving highest performance with least resource.}
\end{tcolorbox}

\begin{tcolorbox}[colframe=gray,left=6pt,right=6pt,boxrule=1.5pt,arc=.5em]\textit{\textbf{Peril 6:} CodeT5's efficiency for generation tasks is in doubt, as the highest resource consumption due to model complexity does not guarantee a consistent better performance on different metrics.}
\end{tcolorbox}

%% file: related.tex
\section{Related Work }
\label{sec:related}
For our related work, we focus on the pre-trained language models, the four applications - code summarization, bug fixing, bug detection, and code search, and related empirical studies of transformers.

\subsection{Pre-trained Language Models}
Different pre-trained language models are developed and demonstrated to have high performance in many NLP tasks~\cite{kenton2019bert,lewis2019bart,liu2019roberta,radford2019language}. 
With the success of pre-trained language models in the NLP domain, researchers have been exploring and applying these models to code-related tasks~\cite{kanade2020learning,liu2020multi,roziere2020unsupervised}. Many pre-trained models for code have been developed. CodeBERT~\cite{feng2020codebert} is one of the earliest models that has been specifically trained for code-related tasks. Subsequently, models like GraphCodeBERT~\cite{guographcodebert} were proposed to improve over CodeBERT by incorporating additional information, such as data flow. Similarly, CodeGPT~\cite{lu2021codexglue} and CodeT5~\cite{wang2021codet5} are built based on GPT and T5 architectures, but pre-trained on a code-related corpus with additional pre-training objectives to better understand code. 
Our experiments are conducted on these three representative pre-trained models for code - CodeBERT, CodeGPT, and CodeT5. Currently, there are many revolutionary large language models being developed and applied to different domains, e.g., GPT-4~\cite{openai2023gpt4} and LLaMA~\cite{touvron2023llama}, which are excluded from our study due to their commercial attribute.

\subsection{Applications}
We study four applications in this paper - code summarization, bug fixing, bug detection, and code search.

\subsubsection{Code Summarization}
Code summarization is one of the most popular tasks in deep learning. 
Fernandes et al.~\cite{fernandes2018structured} combined RNN/Transformers with GGNN. Ahmad et al.~\cite{ahmad2020transformer} applied Transformer to code summarization, and showed the advantage of sequential relative attention over positional encoding. 

\subsubsection{Bug Fixing}
There has been a lot of work in the domain of Transformer-based models that focuses on bug fixing. A majority of this work fine-tunes a pre-trained model for code. For example, CURE~\cite{jiang2021cure} fine-tuned a GPT model to generate patches for buggy code. SPT-Code~\cite{niu2022spt}, which has similar model structure as CodeBERT, also used fine-tuning to perform code refinement.

\subsubsection{Bug Detection}
Many approaches have been explored in the field of bug detection. Over the past decades, developer information has been utilized to predict bugs~\cite{weyuker2007using,meneely2008predicting,pinzger2008can}. With the development of deep learning techniques, Yang et al.~\cite{yang2015deep} leveraged deep learning to generate new features from traditional features using a Deep Belief Network (DBN). Later, Li et al.~\cite{li2017software} generated new features from processing Abstract Syntax Trees (ASTs) of program through Convolutional Neural Network, and combined them with traditional hand-crafted features to perform bug prediction. There are also deep learning algorithms that specialize in bug detection, e.g., DeepJIT~\cite{hoang2019deepjit} and CC2Vec~\cite{hoang2020cc2vec}.

\subsubsection{Code Search}
In deep learning domain for code search, Sachdev et al.~\cite{sachdev2018retrieval} developed the tool NCS to learn embeddings of code without supervision. Gu et al.~\cite{gu2018deep} proposed CODEnn to learn code representations through three encoded individual channels. With the outstanding performance of Transformer-based pre-trained models, many works~\cite{ahmad2021unified,feng2020codebert,mastropaolo2021studying,phan2021cotext} have looked into their application to code search and achieved satisfying results.

\subsection{Empirical Study}
There are some empirical studies on transformers in the literature. For example, Zeng et al.~\cite{zeng2022extensive} has studied the suitability and robustness of different pre-trained models for code understanding and generation tasks. In~\cite{chirkova2021empirical}, Chirkova and Troshin investigated the capabilities of Transformers to leverage syntactic information in different tasks. Wan et al.~\cite{wan2022they} analyzed the interpretability of pre-trained models from different aspects including attention mechanism, word embedding, and syntax tree. Shi et al.~\cite{shi2022we} assessed and improved 4 benchmark datasets widely used in code summarization.

Compared to these previous works, our work additionally reviews the literature, summarizes the most widely studied applications, suggests on developers' needs, contests certain beliefs, concludes on model generalization ability trained on different datasets, and investigates the resource consumption of different models.

%% file: conclusion.tex
\section{Conclusion}
\label{sec:conclusion}

In this empirical study, we comprehensively review the literature on transformer-based pre-trained models used in SE research published from 2017-2022. We focus on the three very representative and widely studied models - CodeBERT, CodeGPT, and CodeT5 - and evaluate their performance on four most popular code-related tasks - Code Summarization, Bug Fixing, Bug Detection, and Code Search. We examine existing literature and developers' needs, contest current beliefs for model architectures, evaluate models' generalization ability to different datasets, and consider the resource consumption of models. Significantly, we also summarize the promises and perils of using transformer-based models for SE research.

%
%
This study highlights several practical issues that need to be addressed in future research on transformer-based models for SE tasks. First, it is crucial to prioritize applications that are most relevant to developers' needs, such as communication and collaboration, deprecated features, module dependencies, and commit issues. Second, the encoder-decoder architecture's optimality for general-purpose coding tasks needs to be re-examined. Third, it is important to carefully select the most suitable model for each specific task. Fourth, the commonly used benchmark datasets should be improved to enhance their generalization ability, making the trained models applicable to other datasets for the same task. Finally, it would be worthwhile to investigate the potential benefits of pruning techniques on trained models to reduce time and space complexity by using a smaller model for inference instead of the original larger model.

 \section{Data Availability} 
Our implementation is
open-source\footnote{https://anonymous.4open.science/r/Transformer-Empirical-SE-63ED} and the datasets we used are publicly available whose access are also showed in the above repository.